\documentclass[a4paper]{article}

\usepackage{amsfonts, amsmath, amssymb, amsbsy, graphicx}
\usepackage[utf8]{inputenc}
\usepackage[english, ngerman]{babel}
\usepackage[T1]{fontenc}
\usepackage{wrapfig}

\def\degr{\hbox{$^\circ$}}

\begin{document}
\selectlanguage{english}

\begin{center}
\textbf{\LARGE A study of photometric errors on two different photographic plate scans}
\end{center}

\begin{center}
\textbf{M. Spasovic$^1$, C. Dersch$^1$,  A. Schrimpf$^1$, P. Kroll$^2$}
\end{center}

\begin{center}
{\it
\noindent $^1$History of Astronomy and Observational Astronomy, Physics Department, Philipps University Marburg \\
$^2$Sonneberg Observatory, Sonneberg, Germany }
\end{center}

\begin{abstract}
A considerable number of photographic plate archives exist world wide and digitization 
 is in progress or already has been finished. Not only different type of scanners were used but also spatial resolution and dynamic range often were limited due to process duration and storage space. The open question is the effect of these limitations on the results.
61 high resolution photographic plates of the Gamma Cyg field from the Bruce astrograph at Landessternwarte Heidelberg--Königstuhl (aperture 40~cm, focal length 200~cm) had been digitized both in Heidelberg and Sonneberg. Both scanners were set to 16 bit dynamic range. The Heidelberg scanner was operated at 2540 dpi resolution, resulting in a scale of 1 arcsec/pixel, while the Sonneberg scanner was operated at 1200 dpi, yielding a scale of 2.1 arsec/pixel.

In the presented study the standard deviation of non--variable star light curves were examined in dependence of brightness and plate coordinates in both series. No evident differences could be found.
A comparison of the analysis of both scan series will be presented.
\\
\\
\noindent Poster presented at the Annual meeting of the German Astronomical Society, Göttingen (2017) 
\end{abstract}

\section{Introduction}

For a comparative study of astrometry and photometry 61 high resolution photographic blue sensitive plates of the Gamma Cyg field from the Bruce astrograph at Landessternwarte Heidelberg-Königstuhl (aperture 40~cm, focal length 200~cm), which have been digitized both in Heidelberg and Sonneberg, were used: (a)  Heidelberg scanner, 16 bit dynamic range, 2540 dpi resolution, resulting scale of 1.03 arcsec/pixel \cite{heidscan}
and (b) Sonneberg scanner, 16 bit dynamic range, 1200 dpi resolution, resulting scale of 2.15. arcsec/pixel \cite{sonn}

\begin{figure*}[htp!]
\centering			

	\includegraphics[width=0.4\textwidth]{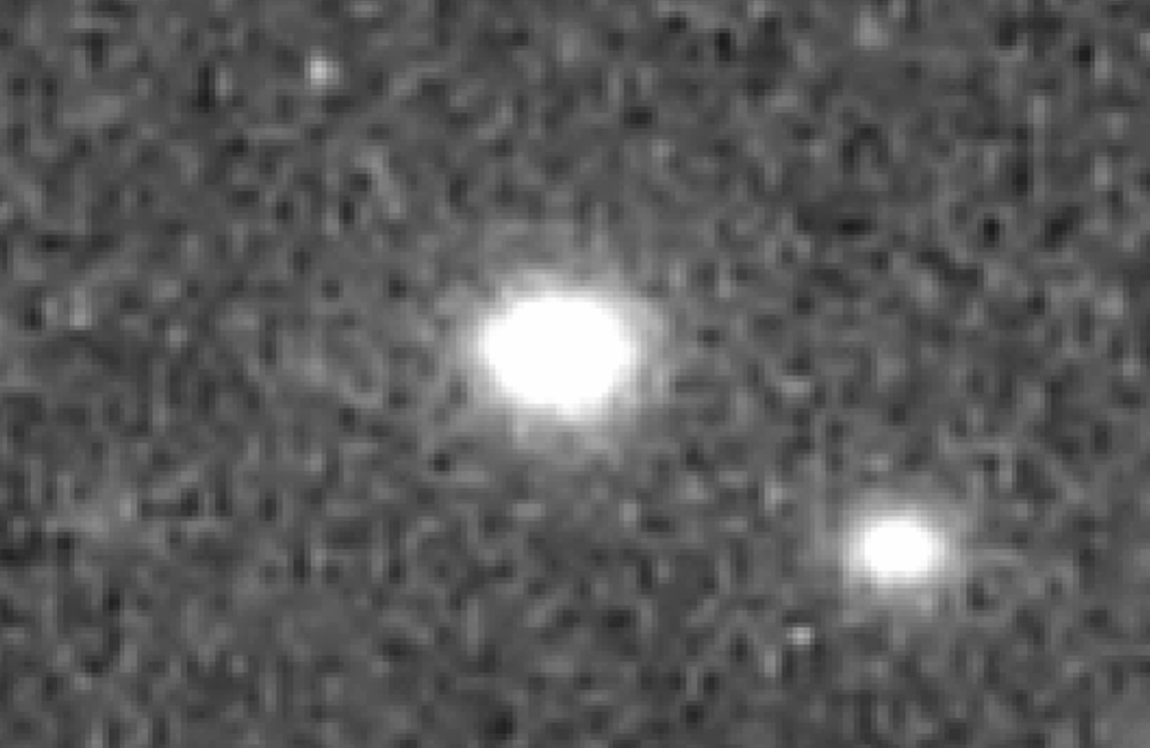}
	\hspace{2em}
	\includegraphics[width=0.4\textwidth]{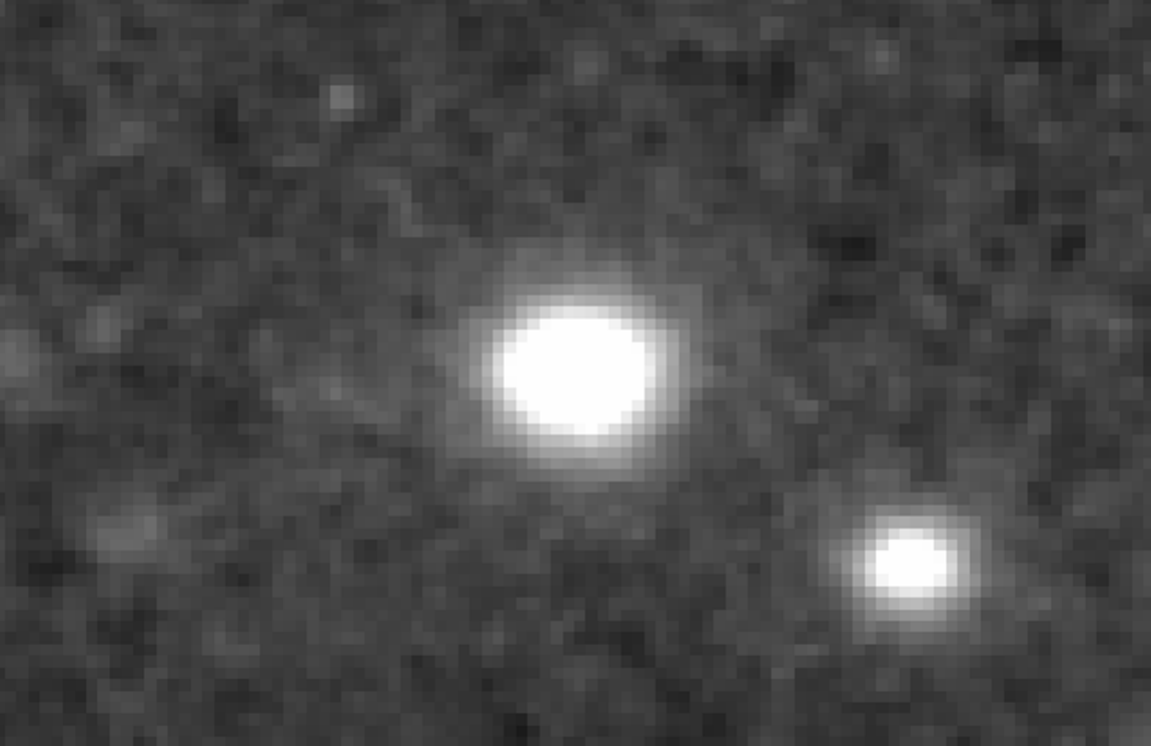}

\caption{Left: Small section of the digitized photographic plate from Bruce Astrograph scanned in Heidelberg. Right: Small section of the digitized photographic plate from Bruce Astrograph scanned in Sonneberg}			
		
\end{figure*}
				    
The difference in resolution between both scans is clearly visible, the question arises as to whether or not this affects the photometry and astrometry.

\section{Astrometry}

The size of the field of theses plates is $6.6\degr \times 8.2\degr$. The astrometric solutions were calculated applying solve-field from the Astrometry.net package \cite{solve}. Using the TAN-SIP WCS correction of higher polynomial order (best results with 5th or 6th polynomial order) we have obtained satisfactory results for calibration and identification of stars with astrometric errors in the range of $\sim$ 5 -- 7 arcsec. By dividing the plates into smaller parts and calculating astrometric solution on each of these sub images as proposed by pyplate \cite{pyplate} astrometric errors about 2 arcsec were determined.		
				
\begin{figure*}[htp!]
	\centering

	\includegraphics[width=\textwidth]{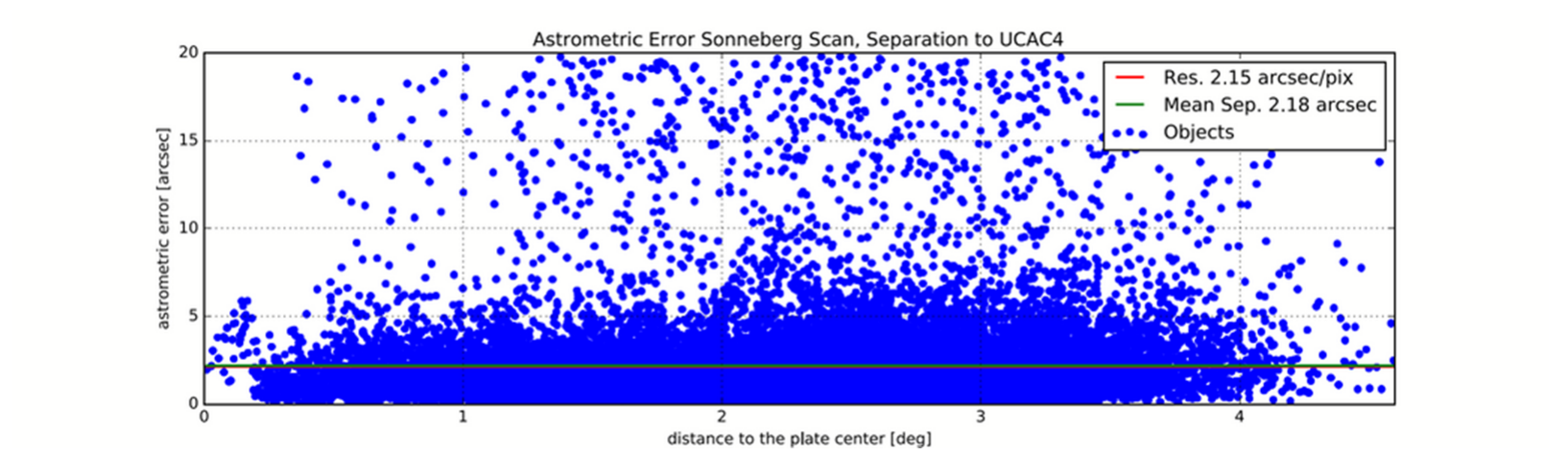}

	\includegraphics[width=\textwidth]{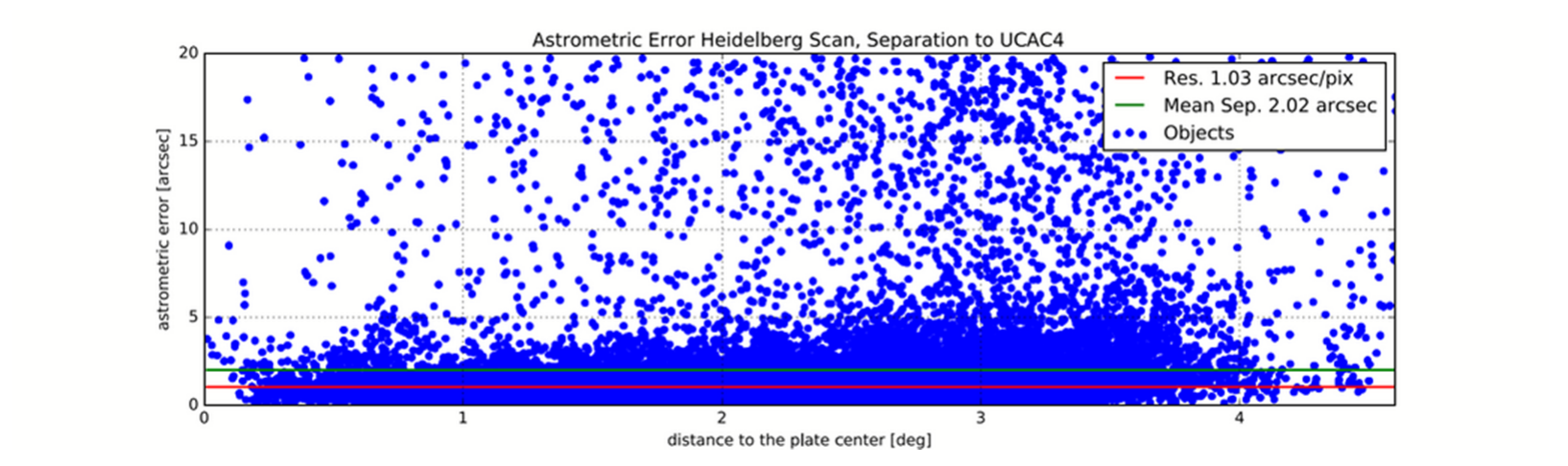}

	\caption{Astrometric errors of stars in a Sonneberg scan (upper part) and in a Heidelberg scan (lower part), separations of stars (blue circles), resolution of the scan (red line) and mean separation to UCAC4 coordinates (green line).}
	\label{astrometry}

\end{figure*}

Calculated separations of positions from identified stars to their positions in the UCAC4 catalog in dependence on the distance from the center of a plate are show in fig. \ref{astrometry}. The Heidelberg scans, due to higher scan resolution, yield slightly better results in astrometry (2.02 arcsec vs 2.18 arcsec).

\section{Photometry}
For source detection the Source Extractor package \cite{sextractor} was used, instrumental magnitudes were computed using the detection threshold as the lowest isophotes. 
For calibration the AAVSO APASS DR9 catalog was used, Only sources with high quality were chosen:  a) stars with B magnitude uncertainty \textless  0.1 mag, and b) errors estimated from standard deviation of several observations.

\begin{figure*}[htp!]
	\centering

	\includegraphics[width=\textwidth]{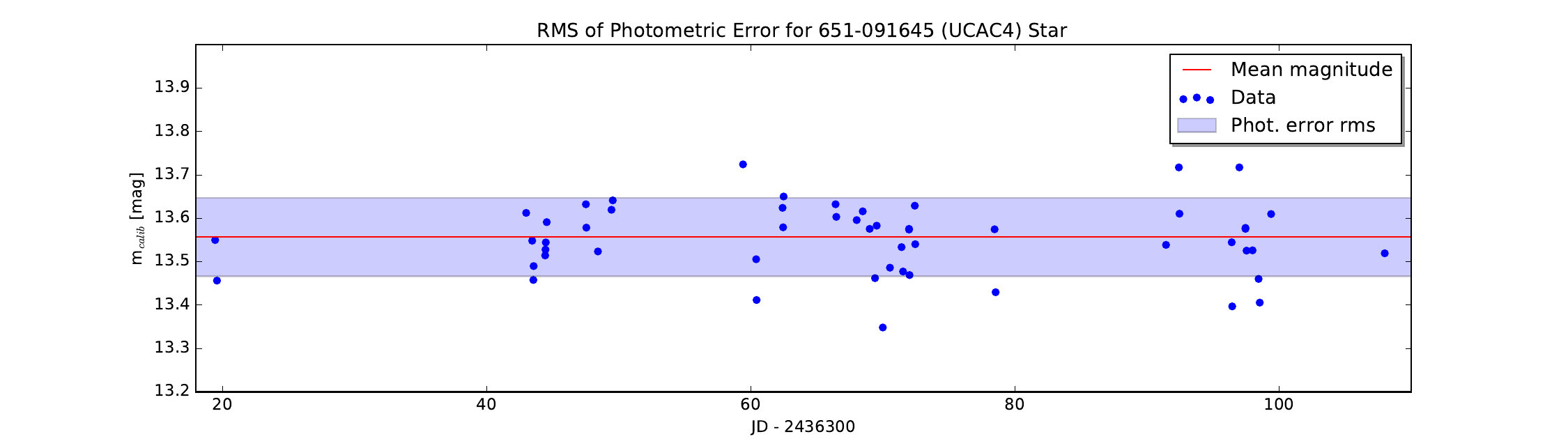}
	\caption{RMS of photometric error of 651-091645 (UCAC4 number), APASS DR9 number 14050429. Catalog magnitude B = 13.38 mag. Calibrated plate mean magnitude m$_{calib}$ = 13.55 mag.}
	\label{photometry}

\end{figure*}	

Photometric errors were analyzed for the stars identified in both scans of the same plates. In fig. \ref{photometry} the results for a constant star throughout the series of plates is shown. For further analysis the scattering of the calibrated magnitude m$_{calib}$ over the time period, i.e.  the root mean square (rms) of photometric error for every single star was calculated.

\section{Photometric Errors}

\begin{figure*}[htp!]
	\centering
	
	\includegraphics[width=0.4\textwidth]{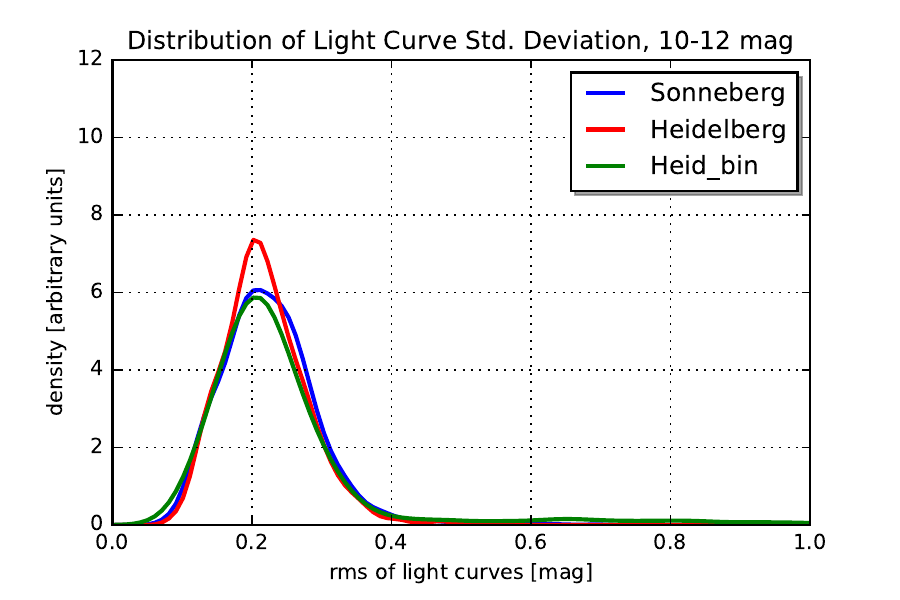}
	\includegraphics[width=0.4\textwidth]{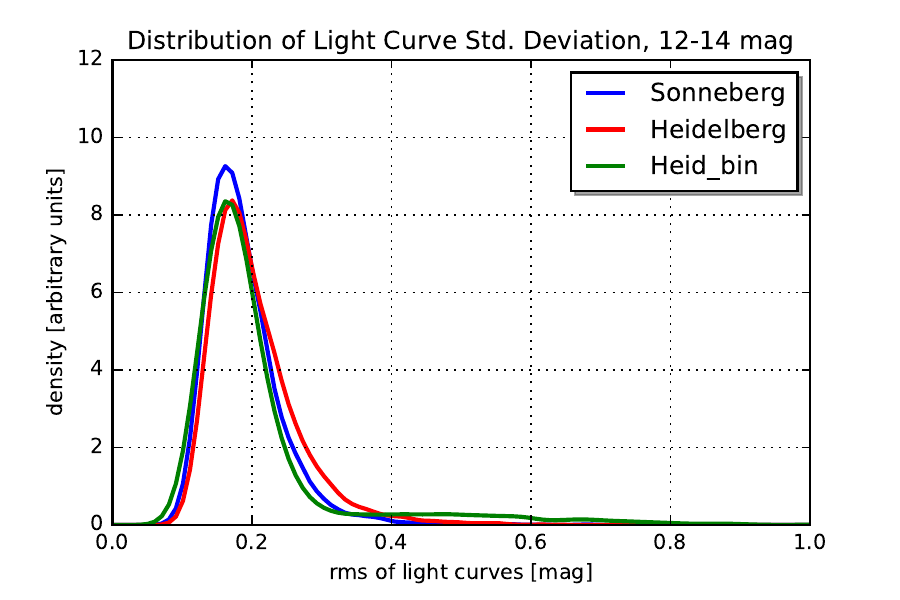}

	\hspace{1em} 
					 					 
	\includegraphics[width=0.4\textwidth]{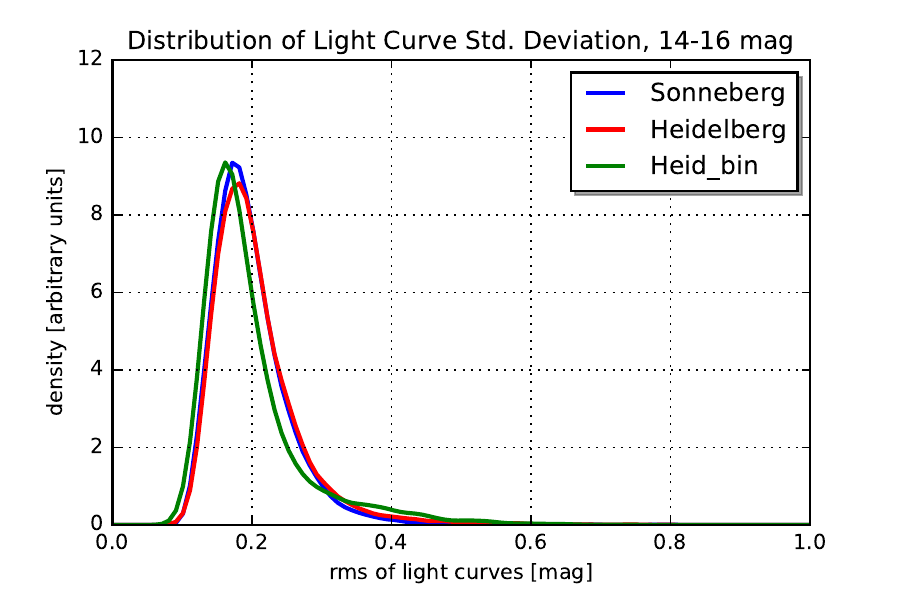}					
	\includegraphics[width=0.4\textwidth]{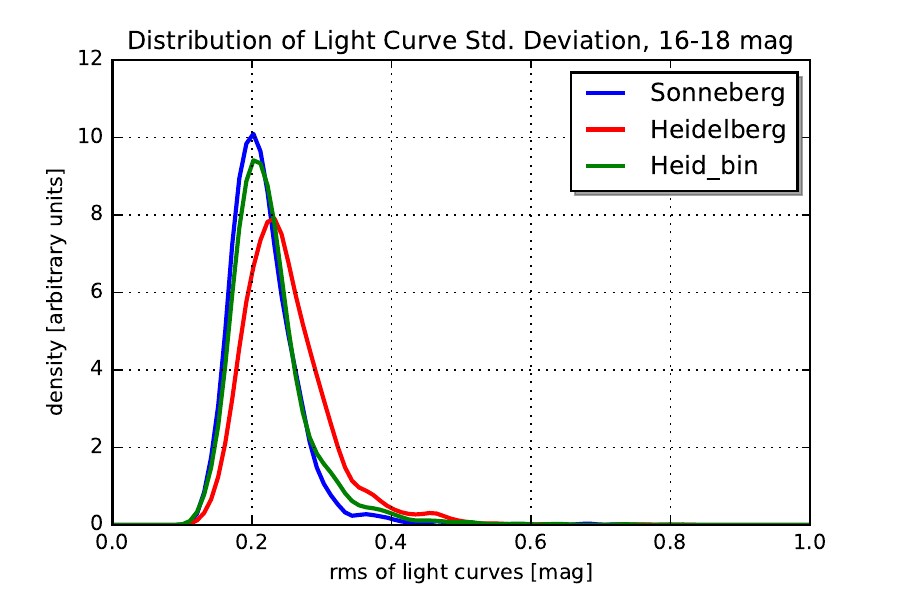}

	\caption{Kernel density estimator of the distributions of photometric rms for Sonneberg and Heidelberg scans in for different magnitude bins.}
	\label{photometric_errors}
						
\end{figure*}

The distribution of photometric errors of all identified stars of both scan series is smoothed with a kernel density estimator. For comparison the photometric errors of 2x2 binned Heidelberg scans, which corresponds roughly to Sonneberg scan resolution, were determined, too (fig. \ref{photometric_errors}.)

The distribution of the brightest stars in the range from 10 to 12 mag is broadened compared to those of the fainter stars. Obviously in these plates stars in this magnitude range are saturated. In the brightness range from 12 to 16 mag the photometric errors of both scan series are almost identical. The Heidelberg scans show a slightly wider error distribution for fainter stars (16 -- 18 mag).

\section{Results}

The most important result of this analysis is that no obvious differences in astrometric and photometric solutions were found. I.e. the higher scanner resolution used in the Heidelberg scans is not necessary for these plates. Surprisingly, the low resolution Sonneberg scans show a better result for fainter stars. And, binning the Heidelberg scans improved the photometry of faint stars. One possible explanation is the use of isophotos to detect grouped pixels of a star above a certain threshold. Higher resolution scans  may lead to isolated signal pixels, which could be missed by isophotal photometry. There are two problems in analyzing astronomical photoplates: (1) They usually cover a large field size than CCD detectors and thus determining a plate solution is possible only by considering field distortions. And (2) the sensitivity of emulsions is non-linear. A usual approach to the second problem is to calculate the calibration in separate annular bins. These two problems should influence the results in the same way in any magnitude range. However, this could not be found.

Further analysis is necessary, especially on the surprising results of the fainter stars. And, the analysis of scans with different scanner resolution of low resolution plates, which exist in much higher numbers, is still pending.

\end{document}